\documentclass[compsoc,conference,a4paper,10pt,times]{IEEEtran}
\usepackage{graphicx}
\usepackage{algorithm}
\usepackage{listings}
\usepackage{gensymb}
\usepackage[utf8]{inputenc}
\DeclareUnicodeCharacter{2212}{-}
\usepackage{mathtools}
\usepackage{algpseudocode}
\usepackage{multirow}
\usepackage{amssymb}
\usepackage{url}
\usepackage{hyperref}
\usepackage{xcolor,colortbl}
\usepackage{caption}    % To create space between listing caption and listing
\usepackage{booktabs}
\usepackage{pifont}
\usepackage{verbatim}
\usepackage{tikz}

\newcommand{\attackName}{Signature Correction }

\begin{document}

\title{\attackName Attack on Dilithium Signature Scheme}

\makeatletter
\newcommand{\linebreakand}{%
  \end{@IEEEauthorhalign}
  \hfill\mbox{}\par
  \mbox{}\hfill\begin{@IEEEauthorhalign}
}
\makeatother

\author{\IEEEauthorblockN{Saad Islam}
\IEEEauthorblockA{%\textit{dept. name of organization (of Aff.)} \\
\textit{Worcester Polytechnic Institute}\\
Worcester, MA, USA \\
sislam@wpi.edu}
\and
\IEEEauthorblockN{Koksal Mus}
\IEEEauthorblockA{%\textit{dept. name of organization (of Aff.)} \\
\textit{Worcester Polytechnic Institute}\\
Worcester, MA, USA \\
kmus@wpi.edu}
\and
\IEEEauthorblockN{Richa Singh}
\IEEEauthorblockA{%\textit{dept. name of organization (of Aff.)} \\
\textit{Worcester Polytechnic Institute}\\
Worcester, MA, USA \\
rsingh7@wpi.edu}
\makeatletter
\linebreakand
\IEEEauthorblockN{Patrick Schaumont}
\IEEEauthorblockA{%\textit{dept. name of organization (of Aff.)} \\
\textit{Worcester Polytechnic Institute}\\
Worcester, MA, USA \\
pschaumont@wpi.edu}
\and
\IEEEauthorblockN{Berk Sunar}
\IEEEauthorblockA{%\textit{dept. name of organization (of Aff.)} \\
\textit{Worcester Polytechnic Institute}\\
Worcester, MA, USA \\
sunar@wpi.edu}

}

\maketitle

\begin{abstract}
Motivated by the rise of quantum computers, existing public-key cryptosystems are expected to be replaced by post-quantum schemes in the next decade in billions of devices. To facilitate the transition, NIST is running a standardization process which is currently in its final Round. Only three digital signature schemes are left in the competition, among which Dilithium and Falcon are the ones based on lattices. Besides security and performance, significant attention has been given to resistance against implementation attacks that target side-channel leakage or fault injection response. Classical fault attacks on signature schemes make use of pairs of faulty and correct signatures to recover the secret key which only works on deterministic schemes. To counter such attacks, Dilithium offers a randomized version which makes each signature unique, even when signing identical messages.

In this work, we introduce a novel \attackName Attack which not only applies to the deterministic version but also to the randomized version of Dilithium and is effective even on constant-time implementations using \texttt{AVX2} instructions. The \attackName Attack exploits the mathematical structure of Dilithium to recover the secret key bits by using faulty signatures and the public-key. It can work for any fault mechanism which can induce single bit-flips. For demonstration, we are using Rowhammer induced faults. Thus, our attack does not require any physical access or special privileges, and hence could be also implemented on shared cloud servers. Using Rowhammer attack, we inject bit flips into the secret key $s_1$ of Dilithium, which results in incorrect signatures being generated by the singing algorithm. Since we can find the correct signature using our \attackName algorithm, we can use the difference between the correct and incorrect signatures to infer the location and value of the flipped bit without needing a correct and faulty pair. To quantify the reduction in the security level, we perform a thorough classical and quantum security analysis of Dilithium and successfully recover 1,851 bits out of 3,072 bits of secret key $s_1$ for security level 2. Fully recovered bits are used to reduce the dimension of the lattice whereas partially recovered coefficients are used to to reduce the norm of the secret key coefficients. Further analysis for both primal and dual attacks shows that the lattice strength against quantum attackers is reduced from $2^{128}$ to $2^{81}$ while the strength against classical attackers is reduced from $2^{141}$ to $2^{89}$. Hence, the \attackName Attack may be employed to achieve a practical attack on Dilithium (security level 2) as proposed in Round 3 of the NIST post-quantum standardization process.

\end{abstract}

% no keywords

% For peer review papers, you can put extra information on the cover
% page as needed:
% \ifCLASSOPTIONpeerreview
% \begin{center} \bfseries EDICS Category: 3-BBND \end{center}
% \fi
%
% For peerreview papers, this IEEEtran command inserts a page break and
% creates the second title. It will be ignored for other modes.
\IEEEpeerreviewmaketitle

\section{Introduction}
 In recent years, quantum computers have made steady progress to the point where they are considered a threat to traditional public-key cryptosystems based on the conjectured hardness of problems such as integer factorization and discrete logarithm. In a landmark result, Shor introduced an algorithm \cite{shor1999polynomial} that can solve the classically conjectured hard problems of factorization and discrete logarithm in polynomial time with the aid of a quantum computer. Symmetric-key systems will also be affected, albeit to a lesser extent. Using Grover's algorithm \cite{grover1996fast} one may recover symmetric keys by searching through the key-space with square-root time complexity. Hence one may overcome Grover, by equivalently doubling key lengths of symmetric schemes and output sizes of hash functions. As Key Exchange Mechanism (KEM) uses public-key schemes to exchange the symmetric keys, there is a need to develop schemes based on quantum-secure hard problems.

To aid the transition to post-quantum cryptography (PQC), the US NIST has announced a PQC standardization process in 2016 \cite{nistcompetition}. The process started with 82 submissions for public-key encryption (PKE), key encapsulation mechanisms (KEM) and digital signatures. 69 schemes were passed into Round 1, 26 were able to get into Round 2 and currently there are seven finalists and eight alternate candidates in Round 3 expected to be completed by the end on 2021. Similar schemes were merged together and some were attacked by the cryptographic community and were taken out of the competition  \cite{ding2019new2, ravi2019number, dachman2020lwe, greuet2020attack, apon2020cryptanalysis, son2019note, son2019revisiting, smith2020rainbow, kales2019forgery}. There are five categories based on the underlying hard problems: lattice-based, code-based, hash-based, isogeny-based and multivariate schemes. 
These schemes offer varying key sizes under varying performance figures, but lattice-based schemes have comparatively compact keys and exhibit better performance. Five out of seven finalists belong to the lattice category; two of the lattice schemes are digital signatures with Dilithium \cite{ducas2018crystals} being one of them. It belongs to the CRYSTALS family having another finalist KYBER which is a KEM. Both are based on the conjectured hard module Learning With Errors (LWE) problem.

The cryptographic community as well as companies have started integrating the finalists from the NIST competition into existing cryptographic libraries like OpenSSL. An open-source project named Open Quantum Safe (OQS) aims to support the development and prototyping of quantum-resistant cryptography \cite{stebila2016post}. PQShield \cite{pqshield} is providing four different products for hardware and firmware for embedded devices, SDK for mobile and server technologies and encryption solution for messaging platforms. Another company, QuSecure \cite{qusecure}, is providing a software solution to protect the data at rest. The transition from classic to post-quantum algorithms is urgently needed to ensure forward secrecy.

According to the status report on the second Round of NIST PQC \cite{alagic2020status}, evaluation is based on three criteria: 1) Security. 2) Cost and performance. 3) Algorithm and implementation characteristics. The third criterion is very important since even if a scheme is mathematically secure, it may succumb to side-channel and fault attacks targeting the implementation. Indeed, in recent years, numerous side-channel attacks e.g. \cite{bruinderink2016flush, pessl2017bliss, bootle2018lwe, ravi2018side, pessl2019more, primas2017single, espitau2017side, d2019timing, park2018side, askeland2021side, karabulutfalcon, kim2020novel} and fault attacks e.g. \cite{mus2020quantumhammer, bruinderink2018differential, ravi2019exploiting, pessl2021fault, kramer2019fault, bindel2016lattice, espitau2016loop, bindel2017special, ravi2019number, genet2018practical} have been demonstrated by the research community on PQC schemes. These include cache attacks, power and EM side-channels, EM and laser injections, clock-glitches and the Rowhammer attack. A major challenge in applying these attacks on PQC schemes, is that PQC schemes have massive key sizes (KBytes) while the attacks can reveal only a few bits per attempt. Yet, even a few revealed key bits may reduce the security strength below the  level specified by the PQC standard. Another challenge for side-channel attacks is that all the finalists have constant-time AVX2 implementations for example they do not have secret dependent branches or other timing variations based upon the secret key. Also, the schemes in Round 3 have withstood more than five years of cryptanalysis by the cryptographic community and the underlying hard problems have been analyzed for decades. For the fault attacks like Differential Fault Attacks (DFA), PQC schemes already have a mitigation by randomizing the nonce values. DFA works on a principle of taking the difference of the correct and faulty pair of output and mathematically recover the secret key. After this mitigation, the same message signed or encrypted twice gives a different signature or ciphertext and the attacker is unable to collect a faulty and correct pair of the same message. The exception is the recently introduced fault attack named QuantumHammer \cite{mus2020quantumhammer} which exploits the faulty signatures to recover secret key bits. However, QuantumHammer works only on LUOV, a multivariate signature scheme eliminated in Round 3. 

In this work, we are the first to demonstrate a fault attack on the randomized version of Dilithium in Round 3, which is also applicable to the deterministic version. Previous fault attacks on Dilithium \cite{bruinderink2018differential, ravi2019exploiting, ravi2019number} are only applicable on the deterministic version of Dilithium in Round 1. DFA requires a pair of faulty and correct signatures which can be collected by signing the same message twice and faulting in the second iteration. To prevent this DFA, Round 2 Dilithium introduced signature randomization by using a different nonce for every signature generation. Our proposed \attackName attack is independent of the nonce and hence applicable to both randomized and deterministic versions of Dilithium. Bruinderink {\em et al.} \cite{bruinderink2018differential} based their analysis on hypothetical faults without experimental confirmation. Ravi {\em et al.} \cite{ravi2019exploiting, ravi2019number} have experimented using EM fault injection on the reference implementation for ARM-Cortex-M4. All of these attacks require physical access to induce the faults. 

We propose \attackName attack on Dilithium and demonstrate it on the constant-time AVX2 implementation using a Rowhammer attack. Our \attackName attack can work with any single fault injection mechanism. We have chosen Rowhammer, because it is a software-only fault attack that can be launched remotely. Also, it has not been mitigated and can be dangerous in cloud scenarios where different users shares the same DRAM \cite{xiao2016one, cojocar2020we}. 

%%%%%%%%%%%%%%%%%%%%%%%%%%%%%%%
\subsection{Our Contribution}
%%%%%%%%%%%%%%%%%%%%%%%%%%%%%%%
We introduce the \attackName Attack on the Dilithium signature scheme which recovers secret key bits using only the faulty signatures and the public key. The attacks works by first inducing bit flips in the signing process, then collecting the faulty signatures and finally recovers the secret key bits while trying to correct the faulty signature using verification algorithm as an oracle. The faults are induced using a practical and software only Rowhammer attack to produce the faulty signatures. In summary, in this work:

\begin{enumerate}
\item We introduce the \attackName Attack on Dilithium signature scheme on both randomized as well as deterministic version. The \attackName Attack only requires faulty signatures and the public key to mathematically locate single bit faults on the secret key and to reveal the exact value of the bit-flip independent of the fault mechanism used. 

\item We practically demonstrate the Rowhammer attack as a fault injection mechanism for \attackName on constant-time AVX2 implementation of Dilithium to generate the faulty signatures. Unlike physical fault mechanisms like EM, laser or clock-glitches, Rowhammer does not require any physical access which permits remote attacks on shared servers and is also applicable through JavaScript.

\item We recover partial secret key of 883 bits out of 3,072 bits for Dilithium security level 2 in about 2 hours of online Rowhammer attack and negligible amount of post-processing.

\item Careful analysis of the encoding of the secret key allows us to increase the number of recovered bits from 883 to 1,522. Additionally, analysis on the positions of the recovered bits reveal an additional 329 bits hence significantly extending the key material. Detailed analysis is given in Section~ \ref{sec:complexity}.  

\item  Further analysis of lattice attacks shows a much reduced security for Dilithium security level 2 below the NIST's requirements, i.e. from $2^{128}$ to $2^{81}$. Hence a partial key material collection and recovery with \attackName Attack followed by a lattice attack may indeed compromise Dilithium level 2 in practice.

\item Our \attackName Attack is applicable to all variants of Dilithium currently in Round 3 including the randomized versions recommended for side-channel and fault attacks.

\item We propose countermeasures to detect and prevent the \attackName attack by temporal and spatial redundancy techniques as well as through Rowhammer mitigations.

\end{enumerate}

\subsection{Outline}
In Section \ref{sec:Background}, we describe a brief Background of Dilithium signature scheme and the Rowhammer attack. We explain our novel \attackName attack in Section \ref{sec:attack}. Section \ref{sec:results} includes the experimental results. Lattice attacks with complexity calculations are explained in Section \ref{sec:complexity}.
In Section \ref{sec:countermeasures}, we propose countermeasures for our attack and Section \ref{sec:conclusion} concludes the work. 

\section{Background}
\label{sec:Background}

We first briefly explain the primitives of the Dilithium scheme. This is followed by an overview of the Rowhammer attack as we are using it as tool to demonstrate our \attackName attack.

\subsection{CRYSTALS - Dilithium}
The Cryptographic Suite for Algebraic Lattices (CRYSTALS) consists of two cryptographic schemes, Kyber \cite{bos2018crystals}, a KEM and Dilithium \cite{ducas2018crystals}, a digital signature algorithm. The suite has been submitted to NIST PQC competition by the Crystals team and both the CRYSTALS are among the Round 3 finalists. These algorithms are based on hard problems over module lattices. We will only talk about Dilithium in this work. The security of Dilithium is based on two problems, namely, Learning With Errors (LWE) problem and SelfTargetMSIS problem. Dilithium is essentially based on Bai-Galbraith scheme proposed by Bai and Galbraith \cite{bai2014improved} in 2014. The design of the scheme is based on ``Fiat-Shamir with Aborts'' \cite{lyubashevsky2009}. Dilithium has three security levels 2, 3 and 5 and also have AES versions instead of SHAKE for performance purposes. We shall just briefly explain the key generation, signing and verification algorithms of Dilithium scheme. We refer the reader to the original specifications for details \cite{ducas2018crystals}.

\subsubsection{Key Generation}
The secret key vectors $s_1$ and $s_2$ of lengths $l$ and $k$ are sampled randomly from a uniform distribution. Each element of these vectors is a polynomial in the ring $R_{q} = \mathbb{Z}_{q}[X]/(X^{n} + 1)$ and the coefficients are of size $\eta$, where $q=2^{23}-2^{13}+1$ and $n=256$. Next, a $k\times l$ matrix \textbf{A} is generated whose entries are also from $R_{q}$ with relatively larger coefficients in range $q$. Then the LWE vector $t$ is computed, part of which is kept secret as $t_{0}$ while the other part $t_{1}$ is made public. The matrix \textbf{A} is also made public while $s_1$ and $s_2$ are kept secret. Dilithium key generation process can be seen in Algorithm \ref{alg:dilithium_keygen} where it outputs $pk$ as public key and $sk$ as secret key. Unlike the Bai-Galbraith scheme, where the whole $t$ was made public, Dilithium just makes $t_1$ public to reduce the size of the public key. The signature size however, is relatively increased by a small factor.

\begin{algorithm}
    \caption{Dilithium Key Generation \cite{ducas2018crystals}}
    \label{alg:dilithium_keygen}
    \begin{algorithmic}[1]
    \State \textbf{Output:} $pk$ - Public Key, $sk$ - Secret Key
    \State $\zeta \gets \{0,1\}^{256}$
    \State $(\rho, \varsigma, K) \in \{0,1\}^{256 \times 3} \gets H(\zeta)$
    \State $(s_1, s_2) \in {S_\eta^l \times S^k_\eta} \gets H(\varsigma)$
    \State $\textbf{A} \in R_q^{k \times l} \gets ExpandA(\rho)$
    \State $t \gets \textbf{A} s_1 + s_2$
    \State $(t_1, t_0) \gets Power2Round_q(t, d)$
    \State $\textit{tr} \in \{0,1\}^{384} \gets CRH(\rho \Vert t_1)$
    \State \textbf{return} $(pk = (\rho, t_1), sk = (\rho, K, \textit{tr}, s_1, s_2, t_0))$

 \end{algorithmic}
\end{algorithm}

\subsubsection{Signature Generation}
Dilithium signing has two modes of operation, deterministic which is the default and randomized, recommended for side-channel and fault attacks scenarios. The nonce $y$ is generated using a seed $\rho'$ which is either deterministic or randomized depending upon the mode of operation. The signature $z$ is generated using the expression $z=y+c\cdot s_{1}$, where $c$ is the challenge vector derived as depicted in Algorithm~\ref{alg:dilithium_sign}. An important part of the signing operation is the rejection sampling which checks if the signature $z$ does not leak any secret information. The rejection sampling loop runs for approximately 4 to 7 times until a secure signature is generated. There is a rejection counter $\kappa$ which is incremented in every loop to generate a different nonce $y$ in each iteration.

\begin{algorithm}
    \caption{Dilithium Signature Generation \cite{ducas2018crystals}}
    \label{alg:dilithium_sign}
    \begin{algorithmic}[1]
    \State \textbf{Input:} $sk$ - Secret Key, $M$ - Message
    \State \textbf{Output:} $\sigma$ - Signature
    \State $\textbf{A} \in R_{q}^{k \times l} \gets ExpandA(\rho)$
    \State $\mu \in \{0, 1\}^{384} \gets CRH(tr \parallel M)$
    \State $\kappa \gets 0, (z, h) \gets \perp$
    \State $\rho' \in \{0, 1\}^{384} \gets CRH(K \parallel \mu)$ (or $\rho' \gets \{0, 1\}^{384}$ randomized)  \label{lst:line:randomized}
    \While{$(z, h) = \perp$}
        \State $y \in S_{\gamma1}^l \gets ExpandMask(\rho', \kappa)$
        \State $w \gets \textbf{A}y$
        \State $w_{1} \gets HighBits_{q}(w, 2\gamma_{2})$
        \State $\tilde{c} \in \{0, 1\}^{256} \gets H(\mu \parallel w_{1})$
        \State $c \in B_{\tau} \gets SampleInBall(\tilde{c})$
        \State $z \gets y+c\cdot s_{1}$   \label{lst:line:z}
        \State $r_{0} \gets LowBits_{q}(w - c\cdot s_{2}, 2\gamma_{2})$
        \If {$\Vert z\Vert \geq \gamma_{1} - \beta$ or $\Vert r_{0}\Vert_{\infty} \geq \gamma_{2} - \beta$} \label{lst:line:rejection_sampling}
            \State $(z, h) \gets \perp$
        \Else
            \State $h \gets MakeHint_{q}(-c\cdot t_{0}, w - c\cdot s_{2} + c\cdot t_{0}, 2\gamma_{2})$
            \If {$\Vert c\cdot t_{0}\Vert_{\infty} \geq \gamma_{2}$ or the \# of 1's in $h>\omega$}
                \State $(z, h) \gets \perp$
            \EndIf
        \EndIf
    \State $\kappa \gets \kappa + l$
    \EndWhile
    \State \textbf{return} $\sigma = (z, h, \tilde{c})$
    \end{algorithmic}
\end{algorithm}

\subsubsection{Signature Verification}

The Dilithium verification algorithm computes the challenge vector $\tilde{c}$ and compares it to the $\tilde{c}$ provided in the signature. Also, it checks the range of coefficients of signature $z$ and the weight of the hint $h$. If all the three conditions are met, the signature is verified, otherwise rejected. The hint $h$ is not kept secret since it is needed by the verifier to makeup for $t_0$. We refer to the Dilithium specification for details \cite{ducas2018crystals}.

\begin{algorithm}
    \caption{Dilithium Signature Verification \cite{ducas2018crystals}}
    \label{alg:dilithium_verify}
    \begin{algorithmic}[1]
    \State \textbf{Input:} $pk$ - Public Key, $M$ - Message, $\sigma$ - Signature
    \State \textbf{Output:} Verify / Reject
    \State $\textbf{A} \in R_{q}^{k \times l} \gets ExpandA(\rho)$
    \State $\mu \in \{0, 1\}^{384} \gets CRH(CRH(\rho \parallel t_{1} \parallel M)$
    \State $c \gets SampleInBall(\tilde{c})$
    \State $w_{1}' \gets UseHint_{q}(h, \textbf{A}z - ct_{1}\cdot 2^{d}, 2\gamma_{2})$
    \State \textbf{return} $[\Vert z\Vert_{\infty} < \gamma_{1} - \beta]$ and $[\tilde{c} = H(\mu \parallel w_{1}')]$ and [\# of 1's in $h \leq \omega$]
    \end{algorithmic}
\end{algorithm}

\subsection{Rowhammer Fault Injection Mechanism}
We are using Rowhammer as a tool to inject faults. We briefly review the concept and operation of the Rowhammer attack, covering memory management, DRAM organization, address translation and applicability on cloud environments.

Every process has its own virtual address space which is divided into virtual pages, typically of size 4 KBytes. Memory Management Unit (MMU) translates the virtual addresses into physical addresses and keeps track in form of page tables. The memory controller integrated in modern processor then translates these physical addresses into channels, ranks and banks inside the DRAM. This DRAM addressing varies from system to system and is not publicly disclosed for Intel CPUs, although the DRAM addressing was reverse engineered for some of the systems by Pessl {\em et al.} in 2016 \cite{pessl2016drama}. Each bank then further consists of rows and columns sharing the same row buffer. A DRAM row consists of 64K cells and a cell is composed of a transistor and a capacitor. Data is stored in these capacitors in form of charge and interpreted as a zero or a one according to predefined threshold levels. As capacitors leak charge over time, there is a refresh mechanism to restore the charge of all the DRAM cells every 64ms.

As the DRAM manufacturers are trying to make memories more compact, these rows of cells are getting physically closer leading to disturbance errors from one DRAM row to another. If one row is accessed repeatedly, it might cause electrical interference with the neighboring row due to insufficient insulation and the cells in the neighboring row may leak faster. If the leakage is faster than the refresh frequency, the cells can not maintain their state, which may lead to bit flips. This is known as the Rowhammer effect which was first introduced by Kim {\em et al.} in 2014 \cite{kim2014flipping}. Using Rowhammer, an attacker with access to a row next to the victim row in DRAM is able to cause bit flips in the victim memory, even when the attacker resides in a process completely separate from the victim process. If the attacker hammers one row which causes bit flips in the neighboring row, it is called single-sided Rowhammer.

After this discovery, Seaborn {\em et al.} \cite{seaborn2015exploiting} introduced the double-sided Rowhammer which is far more effective than the earlier single-sided Rowhammer. In a double-sided Rowhammer, the attacker hammers two rows sandwiching the victim row, leaking the victim cells even faster. Veen {\em et al.} \cite{van2016drammer} in 2016 showed that it is also applicable on mobile platforms. Gruss {\em et al.} \cite{gruss2018another} introduced one-location hammering and achieved root access with opcode flipping in \texttt{sudo} binary in 2018. Gruss {\em et al.} \cite{2016rowhammerjs} and Ridder {\em et al.} \cite{desmash} have shown that Rowhammer can be applied through JavaScript remotely. Tatar {\em et al.} \cite{tatar2018throwhammer} and Lip {\em et al.} \cite{lipp2020nethammer} have proved that it can be executed over the network. Rowhammer is also applicable in cloud environments \cite{xiao2016one, cojocar2020we} and heterogeneous FPGA-CPU platforms \cite{weissman2019jackhammer}. In 2020, Kwong {\em et al.} \cite{kwong2020rambleed} demonstrated that Rowhammer is not just an integrity problem but also a confidentiality problem.

There have been many efforts on Rowhammer detection \cite{irazoqui2016mascat, chiappetta2016real, zhang2016cloudradar, herath2015these, payer2016hexpads, gruss2016flush+, aweke2016anvil, corbet} and neutralization \cite{2016rowhammerjs, van2016drammer, brasser2017can}. Gruss {\em et al.} \cite{gruss2018another} have shown that all of these countermeasures are ineffective. Some countermeasures require hardware modification, bootloader or BIOS update \cite{brasser2017can, trr, kim2014architectural, kim2014flipping, ghasempour2015armor, ibmchipkill} but they are not all implemented. Cojocar {\em et al.} \cite{cojocar2019ecc} in 2019 reverse engineered the ECC memories showing that ECC countermeasure is not secure either. Another hardware countermeasure Target Row Refresh (TRR) has also been recently bypassed by Frigo {\em et al.} \cite{frigo2020trrespass} using many-sided Rowhammer on DDR4 chips. The same work has been extended by Ridder {\em et al.} \cite{desmash} to attack TRR enabled DDR4 chips from JavaScript. They claim that more than 80\% of the DRAM chips in the market are still vulnerable to the Rowhammer attack.

\section{\attackName Attack on Dilithium}
\label{sec:attack}

To the best of our knowledge, there is no published work yet summarizing a fault attack on Dilithium which can work on randomized version of Dilithium. The randomized version randomly generates the nonce for each signing operation, which gives a different signature every time we sign the same message. Hence a standard DFA is not possible in case of randomized Dilithium as the attacker cannot recover a faulty and another correct signature for the same message for the same nonce. Our novel \attackName attack however is independent of the nonce, hence it is applicable to both randomized and deterministic versions of Dilithium.

The \attackName attack exploits the mathematical structure of Dilithium to recover the secret key bits by using just the faulty signatures and the public key. Thus the attack can be executed offline after collecting sufficiently many faulty signatures from an active fault attack. The attack is independent of the concrete fault injection technique. The only requirement is that the faults should be single bit and induced before the signing step \ref{lst:line:z} of Algorithm \ref{alg:dilithium_sign} in secret key $s_1$. First we define the attacker model and then explain the phases of our \attackName attack.

\subsection{Attacker Model}
When multiple tenants in cloud environments reside on the same server, they may share the same DRAM. The Rowhammer attack requires the attacker process and victim process to share a DRAM. The attacker process can then induce bit flips by just reading its own memory repeatedly \cite{xiao2016one, cojocar2020we, weissman2019jackhammer}. Moreover, the DRAM must be vulnerable to Rowhammer attack which means that its memory cells must be susceptible to the hammering effect. Most types of DRAMs have been shown to be vulnerable in \cite{frigo2020trrespass, desmash}. We are not using HugePages for contiguous memory as most of the servers are not configured to use HugePages. We will explain how we detect contiguous memory in Section \ref{sec:attack} as it is required for the double-sided Rowhammer to locate the neighboring rows in a DRAM bank. Also, the attacker has no knowledge of the DRAM mapping which is different for different memory controllers and DRAM configurations. The DRAM mapping maps physical addresses to actual DRAM ranks, banks, rows and columns which can be used by the attacker to co-locate with the victim in the same DRAM bank. In Section \ref{sec:attack}, we will explain how we use the row conflict side-channel for bank co-location. The attacker can induce bit flips in the secret key $s_1$ of Dilithium but she has no control over the position of bit flip within the $s_1$. For security level 2 for example, the size of $s_1$ is 4 KBytes and the attacker has no knowledge of location of the bit flip within this 4 KBytes memory. Also, she has no idea of the value of the flipped bit. The attacker can just induce bit flips from her own process and is able to collect the faulty signatures from the victim. She can only use these faulty signatures along with the pubic parameters to recover the secret key bits. 

\subsection{Phases of the \attackName Attack}
There are three phases in the \attackName Attack. First, we identify vulnerable memory locations called as templating. Then, we perform double-sided Rowhammer attack on the victim in the online phase and collect the faulty signatures. Finally, we post-process the faulty signatures and recover the flipped secret key bits by \attackName algorithm.

\begin{enumerate}

    \item {\bf Templating Phase:} In a pre-processing phase of the Rowhammer attack, the attacker will identify vulnerable memory locations. The victim needs not to be present during this phase.

    \item {\bf Online Phase:} In the online phase, the victim process is forced to map onto the identified vulnerable memory locations from the templating phase. Then the attacker induces bit flips inside the victim process and collects the faulty signatures generated by the victim. 
   
    \item {\bf Post-processing Phase:} In this phase, the attacker uses the faulty signatures and the public key to recover the secret key bits using the \emph{\attackName algorithm}. This phase can be carried out offline and can be parallelized and run on distributed systems for performance.
    
\end{enumerate}

We will first explain our novel \attackName algorithm. Next, we describe the templating and online phase of Rowhammer to practically demonstrate the fault injection.

\subsection{\attackName Algorithm for Dilithium}
\label{sec:attack_algo}

\attackName is a way to recover the flipped secret key bits using faulty signatures. Since challenge $c$ is public, the generated error in the signature can be used to find the position of the flipped bit in the secret key. The error in the faulty signature can be some certain multiples of $c$. Therefore, if we somehow correct the faulty signature, we are able to find the position of the bit-flip. The main idea of \attackName is to find the faulted bit in the secret key by the process of correcting the faulty signature by checking it using signature verification algorithm. The main difference between the standard DFA and \attackName Attack is that the attacker does not need to know the original signature. Finding the position of the flipped bit by the fault is different for every algorithm. In Algorithm \ref{alg:attack}, we explain it specifically for Dilithium.  

\subsubsection{How to trace back to the flipped bit using a faulty signature}

$s_1$ is defined in $S^l_\eta$ in Algorithm \ref{alg:dilithium_keygen} step 4. Let $s_1=(s_1^{(1)},\cdots ,s_1^{(l)})$ in vector form where $s_1^{(i)}=\Sigma_{j=0}^{n-1} a_j^{(i)}x^j$ and $-\eta\leq a^{(i)}_j\leq \eta$, $1\leq i\leq l$ and $0\leq j\leq n-1$. In Algorithm \ref{alg:dilithium_sign} step 13, signature is generated by $z=y+c\cdot s_1$ where $c=\Sigma_{j=0}^{n-1} c_jx^j$ is a constant challenge vector. If one bit in $s_1$ is flipped before the signature generation, it faults the output signature $\bar{z}=y+c\cdot \bar{s_1}$. Then, the difference of the faulty and original signatures is $\Delta z=z+\bar{z}=c\cdot (s_1+\bar{s_1})=c\cdot \Delta s_1$. Since just one bit is flipped in $s_1$, $\Delta z$ has just one non-zero component which is $c_t\bar{a}_r^{(i)}x^{t+r}$, where $\bar{a}_r^{(i)}$ is the one bit difference, $x^r$ is the position of the flip in $s_1^{i}$ and $c_t$ is the relevant component of the flipped bit in $c$. Note that, because of $x^r$ term, $c$ shifts to the right $r$ times. Additionally, $\bar{a}_r$ is a power of 2 since it is the 1-bit difference.

For instance, if the flip is in the first coefficient of $s_{1}$, the changes in $z$ appear at the same indices at which $c$ is non-zero. If it is in the second coefficient of $s_{1}$, the changes appear at the non-zero indexes of one bit shifted version of $c$ and so on. This observation makes it possible to trace back to the faulty bit by just using the faulty signature and the public key. We can not only locate the position of the bit flip but also the value of the flipped bit because both have a unique effect on the error.

To recover the secret key bit by just using the faulty signature $\sigma'$, first we unpack the faulty signature to get the unpacked faulty signature $z'$ and the challenge information $\tilde{c}$. Next we sample $\tilde{c}$ to get the challenge vector $c$ and copy it to a temporary variable $\overline{c}$ as we will need to modify it. The idea is to add all $n$ shifted versions of $c$ in the faulty signature $z'$ one by one and try to correct the faulty signature. We can verify the correctness using the Dilithium verification, Algorithm \ref{alg:dilithium_verify}, as an oracle. When the signature with the attempted correction verifies, we can tell that this is the index of the flipped coefficient. We can also tell the value of the flipped bit by trying both addition and subtraction of the shifted versions of $c$. We need to repeat this step for all of the $L$ elements of $s_{1}$ to trace the flipped bit for any of the elements of $s_{1}$.

This procedure works if the bit flip occurs in the LSB of the coefficients. If the flip is the second or third LSB, we need to add a $multiplier$, which is $2^{bit\_index}$. This multiplier is first multiplied with the shifted version of the challenge vector $\overline{c}$ and then added to the faulty signature $z'$. In the Dilithium implementation, the coefficients of $s_{1}$ are stored as \texttt{int32\_t}, but the values of the coefficients range up to four bits depending upon the security level. Hence, we need to check up to three or four bits, we call this number as $B$. The algorithm however is capable of going further but there is no useful information on the MSB side as the remaining bits are the same as last useful LSB. So, we need to keep modifying the public challenge $\overline{c}$, multiply it with the $multiplier$, add it to the faulty signature $z'$ and verify to see if the signature is corrected using the verification oracle. If the signature is correct, the algorithm returns the recovered bit of secret key $s_{1}$ as output. The algorithm needs at most $2 \times B \times L \times n$ number of verification to recover one bit of secret key. In practice however, the code breaks earlier upon finding the location. Algorithm \ref{alg:attack} summarizes our attack.

\begin{algorithm}
    \caption{Novel \attackName Algorithm for Dilithium}
    \label{alg:attack}
    \begin{algorithmic}[1]
    \State \textbf{Input:} $\sigma'$ - Faulty Signature, $M$ - Message, $pk$ - Public Key
    \State \textbf{Output:} (row, col, bit\_index, value) - Recovered secret key bit
    \State $(z', h, \tilde{c}) \gets unpack(\sigma')$ 
    \State $c \gets SampleInBall(\tilde{c})$
	\State $\overline{c} \gets c$
    \For{$bit\_index$ \textbf{from} $1$ \textbf{to} 32}
        \State $multiplier \gets 2^{bit\_index-1}$
		\For {$row$ \textbf{from} $1$ \textbf{to} L} 
		    \For {$col$ \textbf{from} $1$ \textbf{to} N}
		        \State  $\overline{z}[row] \gets z'[row] + multiplier \times \overline{c}$
		        \State  $\overline{\sigma} \gets pack(\overline{z}, h, c)$ 
			    \If {$sig\_verify(pk, M, \overline{\sigma}) = true$} 
                    \State \textbf{return} $(row, col, bit\_index, 1)$
                \Else
                    \State $\overline{c} \gets circ\_shift\_right(\overline{c})$
                \EndIf 
            \EndFor
            \For {$col$ \textbf{from} $1$ \textbf{to} N} 
		        \State  $\overline{z}[row] \gets z'[row] - multiplier \times \overline{c}$
		        \State  $\overline{\sigma} \gets pack(\overline{z}, h, c)$
			    \If {$sig\_verify(pk, M, \overline{\sigma}) = true$} 
                    \State \textbf{return} $(row, col, bit\_index, 0)$
                \Else
                    \State $\overline{c} \gets circ\_shift\_right(\overline{c})$
                \EndIf 
            \EndFor
		\EndFor
	\EndFor
	
    \end{algorithmic}
\end{algorithm}

\subsection{Templating Phase}
\attackName attack needs a fault mechanism which can provide faulty signatures. We are practically inducing faults using Rowhammer, a software-only technique which does not require any physical access to the target machine. Recent research has shown that it can be applied over the network \cite{tatar2018throwhammer, lipp2020nethammer} and even remotely through JavaScript \cite{2016rowhammerjs, desmash}. There is no effective countermeasure to prevent Rowhammer completely in DRAM chips so far. Recent research has demonstrated that it is possible to apply Rowhammer even on DDR4 memories with TRR \cite{frigo2020trrespass} mitigation as well as on ECC memories \cite{cojocar2019ecc}. Templating phase involves three steps: contiguous memory detection, bank co-location and double-sided hammering.

\subsubsection{Contiguous Memory Detection}
For a double-sided Rowhammer, the attacker needs to allocate the rows exactly one above and one below around the victim in the actual DRAM. For this purpose, contiguous memory is a requirement for double-sided Rowhammer. It can be achieved using Huge-pages but that requires special configuration and privileges. We achieve contiguous memory detection using \textsc{Spoiler} \cite{islam2019spoiler} from normal user space without any special privileges. When the spoiler peaks become equally distant apart, the physical addresses become contiguous. Figure \ref{fig:cont_mem} shows the frame numbers of memory pages inside a buffer. We can see the contiguous memory where the frame numbers are linearly increasing. A detailed description of this approach can be found in \cite{islam2019spoiler}.

\begin{figure}[t]
  \centering
  \includegraphics[width=\linewidth]{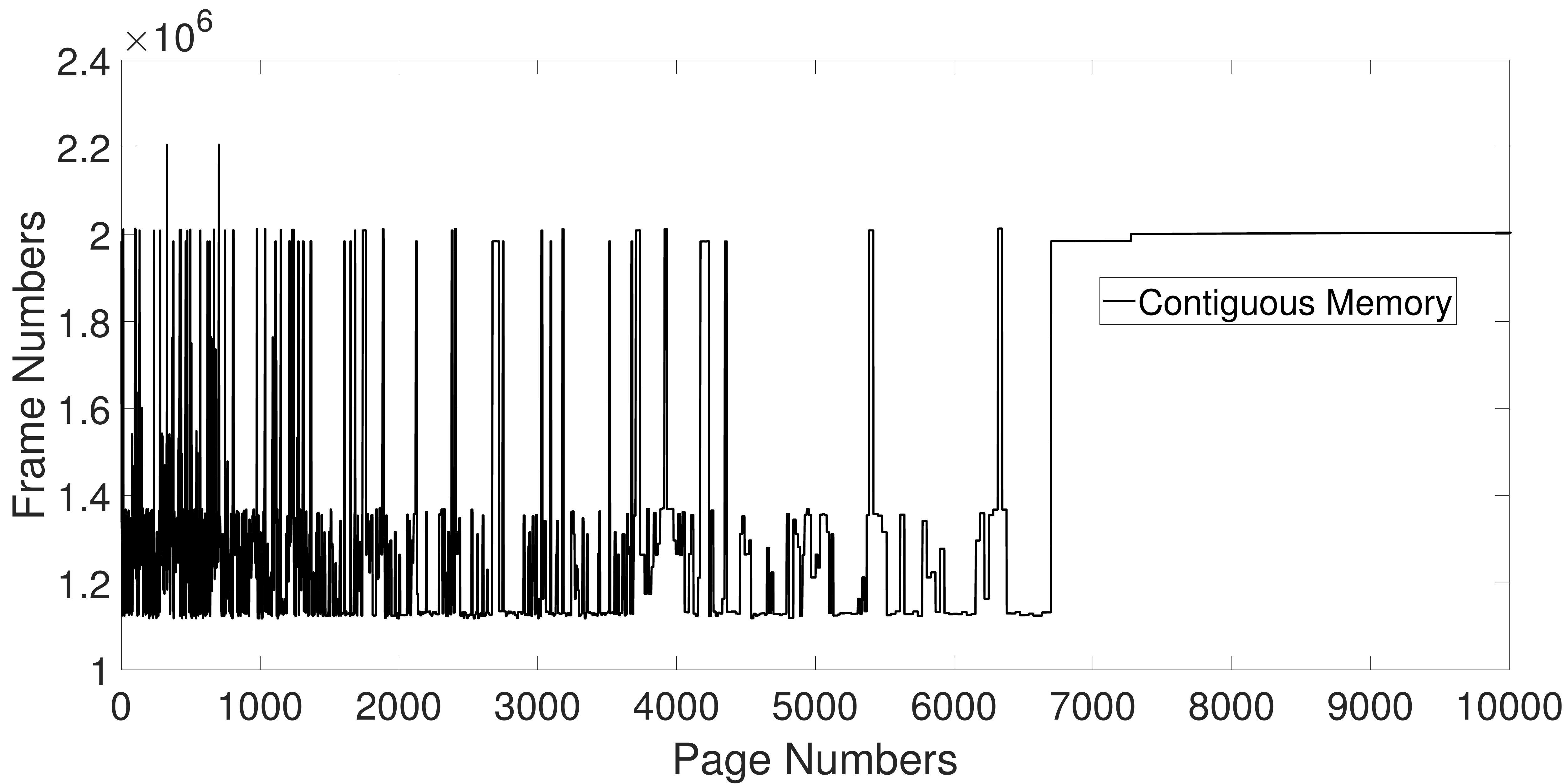}
  \caption{Contiguous memory detection. x-axis shows the page numbers of the allocated memory buffer, each page being 4 KBytes. On y-axis are the frame numbers of these pages in integer form. The straight line shows a linear increase in frame numbers; it is not a horizontal line.}
  \label{fig:cont_mem}
\end{figure}

\subsubsection{Bank Co-location}
A DRAM is organized in banks and every bank has a row buffer. Rowhammer attack works when both the attacker and the victim are sharing the same bank. To find the virtual addresses mapping to the same bank, we use a side-channel which is based on the row conflict. When two addresses from the same bank are accessed, it takes longer as compared to the accesses from different banks. This is because one row loaded inside the row buffer needs to be written back to its original position before loading another row. The CPU cycles taken for accessing one address and the remaining are shown in Figure \ref{fig:row_conflict}. Depending on the maximum values of the peaks, we can set a threshold to extract the addresses mapped to the same bank.

\begin{figure}[t]
  \centering
  \includegraphics[width=\linewidth]{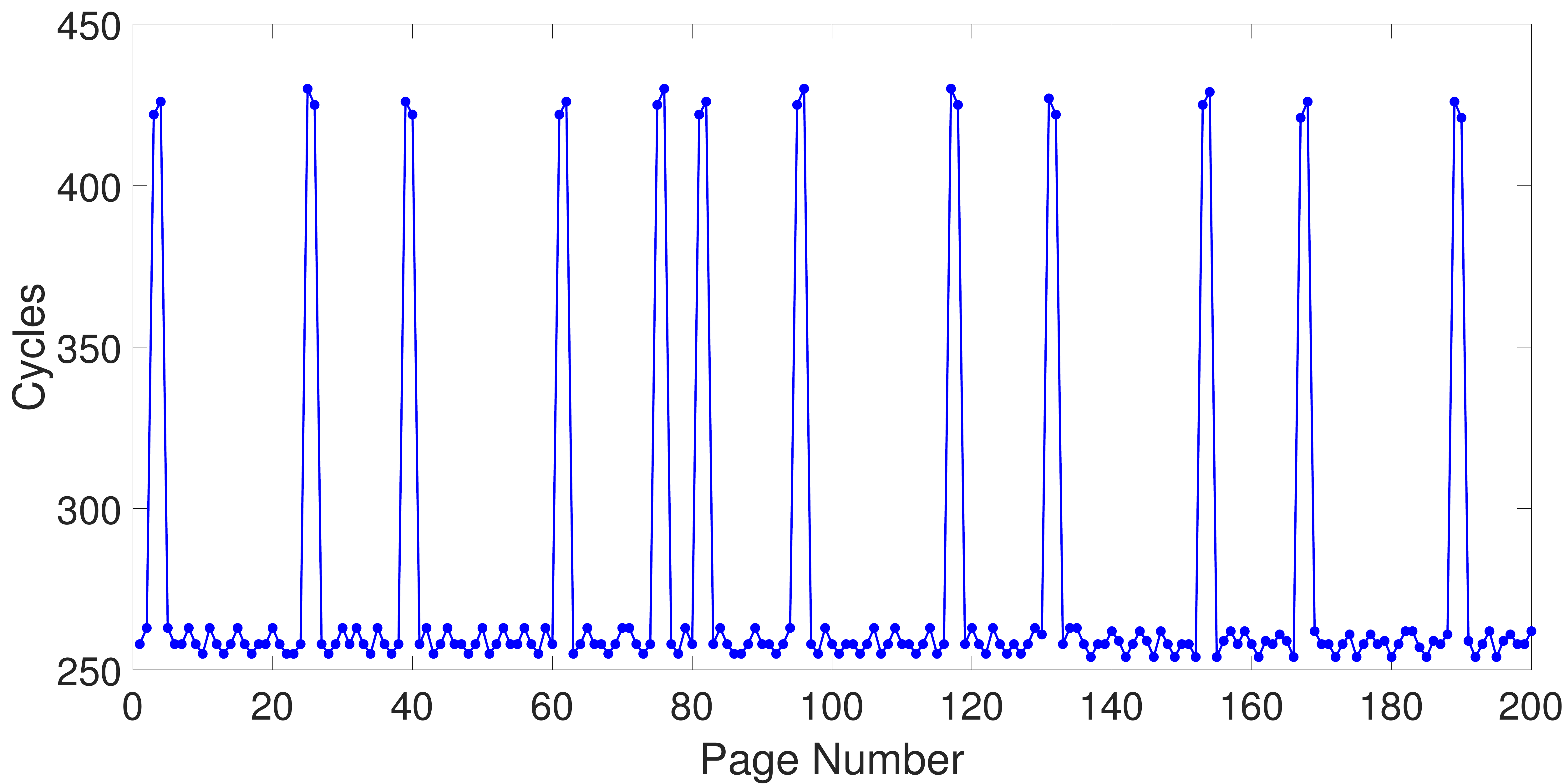}
  \caption{When two DRAM rows are accessed which reside in the same bank, we get a peak due to the row conflict. A threshold can be set to separate these rows using this side-channel information. In our experiments we have set the \texttt{THRESHOLD\_ROW\_CONFLICT} value as 380 cycles.}
  \label{fig:row_conflict}
\end{figure}

\subsubsection{Double-sided Hammering}
Once we identify the contiguous memory within a bank, we can start start taking three rows at a time from this memory and apply double-sided Rowhammer on them. We hammer the top and bottom row and expect the bit flips in the middle row. In our experiments, we have kept the number of hammers equal to $10^6$. While keeping the record of the vulnerable rows, we keep moving onto the next three rows until for our identified contiguous memory. We have used the typical Rowhammer instruction sequences without the \texttt{mfence} as shown in Listing \ref{lst:rowhammer_mfence}. Without the \texttt{mfence}, the number of bit flips are more as compared to with \texttt{mfence}. This is because the DRAM accesses become faster which results in quicker leakage of the charge stored in the memory cells. The number of flips with and without \texttt{mfence} are compared in Figure \ref{fig:flips_vs_hammer}. The number of CPU cycles and the time taken by one Rowhammer instruction sequence is given in Table~\ref{tab:resolution}.

\begin{lstlisting}[float, language={[x86masm]Assembler}, frame=tb, tabsize=2, basicstyle=\footnotesize, caption=Typical Rowhammer instruction sequence \cite{cojocar2020we}, label={lst:rowhammer_mfence}]
loop:
    movzx rax, BYTE PTR [rcx]
    movzx rax, BYTE PTR [rdx]
    clflush BYTE PTR [rcx]
    clflush BYTE PTR [rdx]
    mfence
    jmp loop
\end{lstlisting}

\begin{figure}[t]
  \centering
  \includegraphics[width=\linewidth]{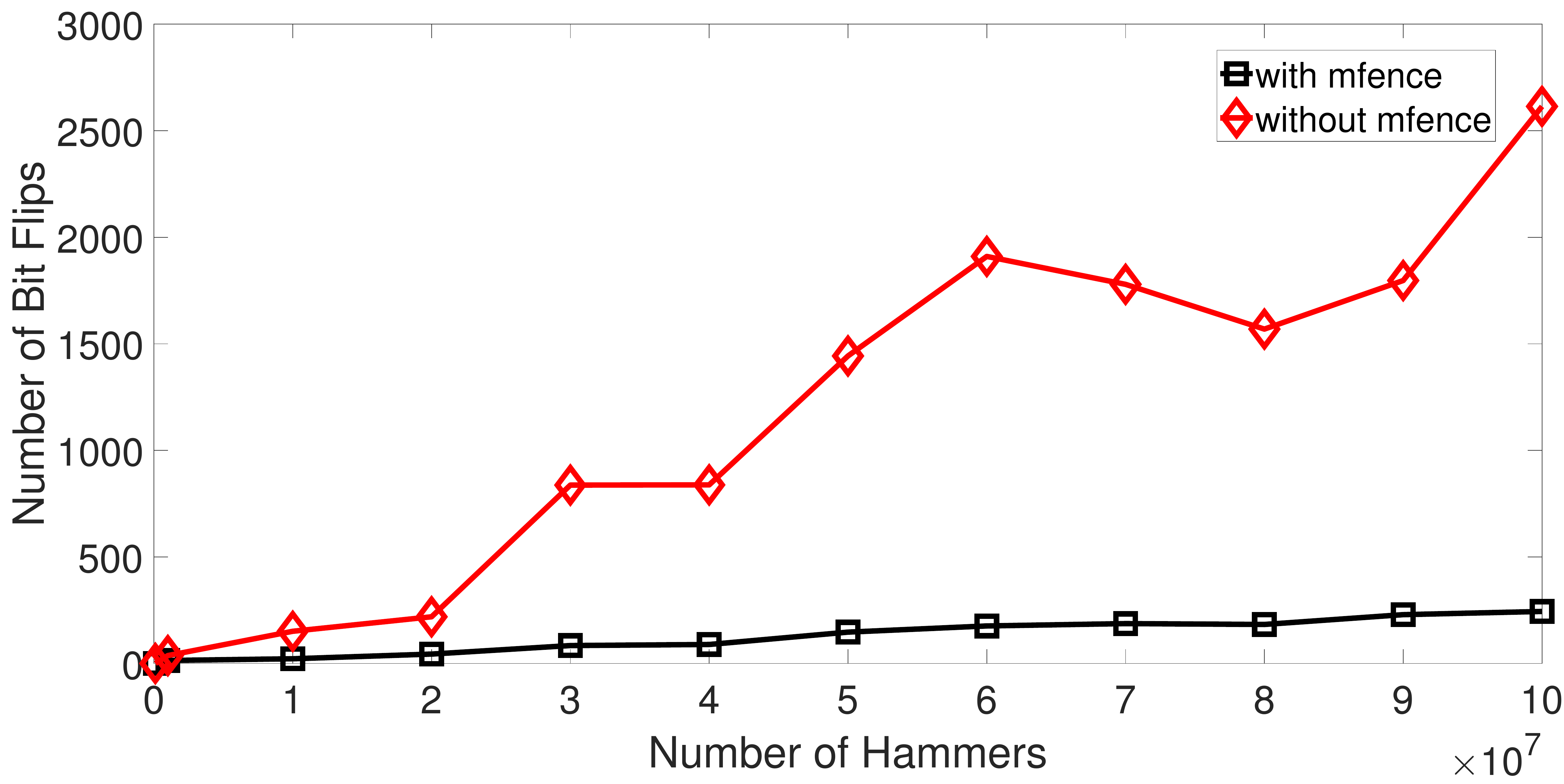}
  \caption{The number of bit flips in 1 MByte of memory in a DRAM bank out of the 8 MBytes contiguous chunk spread across 8 banks as a function of the number of hammers. The number of bit flips increases with the number of hammers and without mfence sequence gives much more bit flips. Approximately 0.03\% of the DRAM cells are found to be vulnerable to Rowhammer attack on the DRAM model we profiled.}
  \label{fig:flips_vs_hammer}
\end{figure}

\begin{table}[t]
\centering
  \caption{CPU cycles and time taken by a typical Rowhammer instruction sequence on our platform.}
  \label{tab:resolution}
  \begin{tabular}{ccc}
   \hline
    Instruction Sequence (mV)&CPU Cycles&Time ($\mu s$)\\
     \hline
    With mfence & 635 & 0.18\\
    Without mfence & 480 & 0.14\\
     \hline
\end{tabular}
\end{table}

\subsection{Online Phase}

As the Rowhammer attack is highly reproducible, we first place the victim into our target vulnerable location inside the memory and repeat the double-sided Rowhammer attack by hammering the neighboring addresses. This induces the bit flips in the actual victim, and faulty signatures are produced by the victim in response. The online phase consists of two steps, first is the victim placement and the second is the double-sided Rowhammer.

\subsubsection{Victim Placement}
\label{sec:victim_placement}
Once the attacker finds vulnerable DRAM rows, it frees the row using \texttt{munmap}. Now it can either wait for the victim page to take that space in the memory or use standard techniques like spraying \cite{2016rowhammerjs, seaborn2015exploiting, xiao2016one}, grooming \cite{van2016drammer} or memory waylaying \cite{gruss2018another, kwong2020rambleed, xu2019memway} to force the victim to come at the target address. We achieve this by repeatedly mapping the secret key $s_{1}$ of the victim until it lands to the target page as shown in Figure \ref{fig:double-sided}. The physical addresses are checked using the \texttt{pagemap} file.

\begin{figure}[t]
  \centering
  \includegraphics[width=0.6\linewidth]{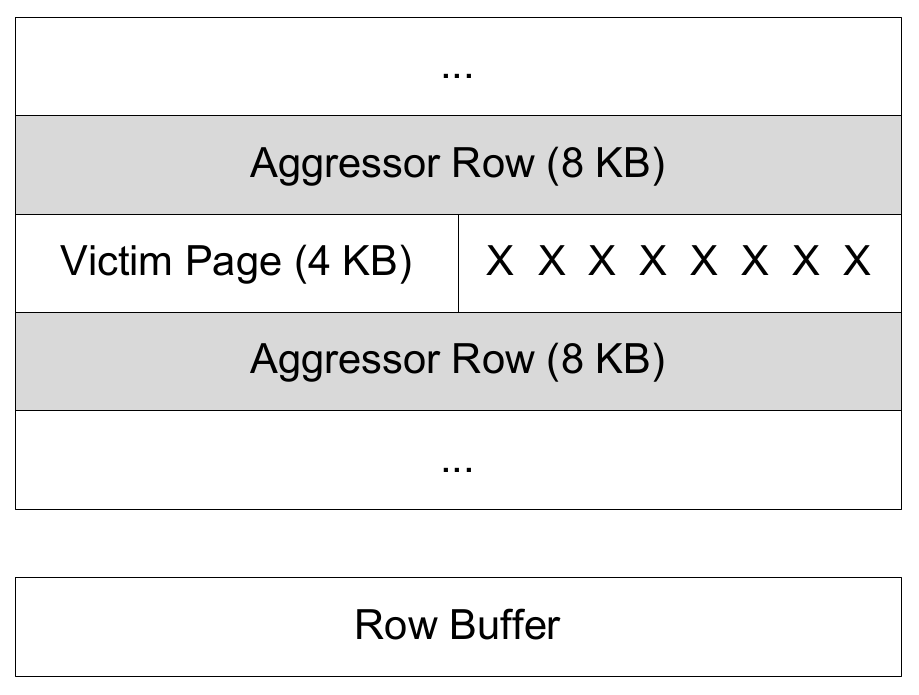}
  \caption{Victim placement and double-sided Rowhammer. To flip the bits from $1\rightarrow0$ inside the victim page, the attacker rows are needed to be filled with all zeros and for $0\rightarrow1$ flips, the attacker rows must be filled with all ones. Empirically, cells which flip both ways are very rare. Hence, a $0\rightarrow1$ flip may not happen in a $1\rightarrow0$ bad cell and vice versa.}
  \label{fig:double-sided}
\end{figure}

\subsubsection{Double-sided hammering}
When the victim is mapped to the attacker's desired vulnerable memory location, the attacker can now apply the double-sided Rowhammer again. While the victim is signing the messages, the attacker now hammers the same rows which she found in the offline phase but this time it flips the bits in the victim process. This is because of the fact that Rowhammer effect is highly reproducible which means if you have found the vulnerable cells once, their values can be flipped again later. Finally, the victim starts producing the faulty signatures due the bit flips in the secret key which are collected by the attacker.

When a bit is flipped on the MSB side of $s_{1}$, it is likely that the rejection sampling condition in step \ref{lst:line:rejection_sampling} of Algorithm \ref{alg:dilithium_sign} repeatedly becomes true or takes too many iterations to output a faulty signature. This can create a \emph{denial of service} scenario and can cause the victim to stuck in a loop and never output a signature unless the victim is moved to another memory location in the DRAM, making our attack harder. To counter this situation, we have set a limit on $\kappa$ in Algorithm \ref{alg:dilithium_sign} to prevent the victim from going into an infinite loop. However, if there is a side-channel attack running in parallel is collecting side information, this scenario can be useful as the nonce $y$ is changing in each iteration.

\section{Experimental Results}
\label{sec:results}

In this Section, first we mention our Rowhammer experimental setup and then mention the results of our \attackName attack experiments\footnote{The source code for \attackName Attack is made available at \url{http://github.com/VernamLab/SignatureCorrection}.}.

\subsection{Experimental Setup}
All the Rowhammer experiments are performed on a Haswell system with Intel(R) Core(TM) i7-4770 CPU @ 3.4GHz with 2 GBytes Samsung DDR3 part number M378B5773DH0-CH9. We have used Haswell because the AVX2 support start from Haswell and it also supports DDR3 memories. Our underlying operating system is Ubuntu 16.04 LTS.

We have performed all the post-processing on a Skylake system with Intel(R) Core(TM) i5-6440HQ CPU @ 2.60GHz having 8 GBytes DDR4 memory running Ubuntu 16.04 LTS using only a single core. The post-processing performance can be improved using multicore, GPUs or distributed computing.

\subsection{Key recovery with \attackName Attack}
\label{sec:key_recovery}
We have successfully applied Rowhammer on $s_1$ of size $1024 \times 32$ bits for the AVX2 implementation of the Dilithium security level 2. After collecting 6,853 single-bit faulty signatures in 2.19 hours of online Rowhammer attack, we have recovered 3,735 unique bits of secret key $s_{1}$ using our \attackName algorithm as shown in Figure~\ref{fig:recovered_bits}. Note that, the faults we can inject are far from uniform. In fact, there are locations that are unflippable. The spatial bias is highly dependent on the technology of the DRAM. In our target DRAM (M378B5773DH0-CH9), we observed heavy spatial correlations (dark vertical stripes in Figure~\ref{fig:recovered_bits}). Also rejection sampling prevents faulty signatures with flips at higher locations to be released. Hence, even if we force $s_1$ to relocate in memory as explained in Section \ref{sec:victim_placement}, this does not allow recovery of all $s_1$ bits. We start recovering the same key bits and while others that wander through unflippable locations are never recovered. Therefore, we stop the online phase and do post-processing after all flippy locations are recovered. Among the 3,735 recovered bits, 2,454 are the $0$'s (green pixels) and 1,281 are the $1$'s (red pixels). Each sub-figure represents an element (polynomial) of $s_{1}$ up to $l = 4$ for Dilithium security level 2. Each polynomial has 256 coefficients on y-axis and 32 bits per coefficient on the x-axis. Every faulty signature gives one bit of secret key. The difference of 3,118 bits is because of the repetition of the faults at the same memory location as the attacker has no control over the locations within the $s_{1}$. 883 out of these 3,735 bits reside in the first three LSBs which should contain the actual key information. The rest of the bits from bit 4 to bit 32 are redundant, same as bit 3.  

However, as the remaining bits from bit 4 to bit 32 are all same as bit 3 for each coefficient, if any of the bits are recovered from this region, we can consider it a bit recovery for LSB 3. This increases our useful bit recovery number significantly from 883 to 1,522 bits. Finally, we can say that by analyzing the positions of recovered bits in the coefficient, we can increase the number of recovered bits from 1,522 to 1,851, see Section~\ref{sec:BitToCoeff} and Section~\ref{sec:BitToNorm} for details. As a summary, we have successfully recovered 1,851 bits out of the total 3,072 bits of $s_{1}$, 3-bits each of 1024 coefficients. The results and distribution of recovered bits up to the secret key coefficients is provided in Table \ref{recoveredbits}.

\begin{table}[t]
\centering
  \caption{Post computation times for \attackName attack on a single CPU. These offline computations can be performed on a distributed system or GPUs for performance improvement.}
  
  \label{tab:attack_performance}
  \begin{tabular}{ccc}
    \hline
    AVX2 & Average CPU Cycles & Time (Sec)\\
    Implementations & (1 Verification) & (1 \attackName)\\
    \hline
    \hline
    dilithium2 & 36595 & 0.094\\
    dilithium3 & 70397 & 0.267\\
    dilithium5 & 67719 & 0.263\\
    dilithium2-AES & 28901 & 0.071\\
    dilithium3-AES & 47614 & 0.177\\
    dilithium5-AES & 49479 & 0.200\\
    \hline
\end{tabular}
\end{table}

Table~\ref{tab:attack_performance} shows the offline computation time needed to trace one bit of secret key for all the variants of Dilithium. These timings are for the worst case scenario of $2 \times B \times L \times n$ verification as explained in Section \ref{sec:attack}. The search is however stopped earlier once a bit is located. We have computed the post-computation times for all variants but demonstrated the Rowhammer attack on only Dilithium security level 2. However our \attackName attack is applicable to all variants, modes and security levels of Dilithium Round 3, where modes are randomized and deterministic, variants are SHAKE and AES and the security levels 2, 3 and 5.

\begin{figure*}[t]
  \centering
  \includegraphics[width=1.07\textwidth]{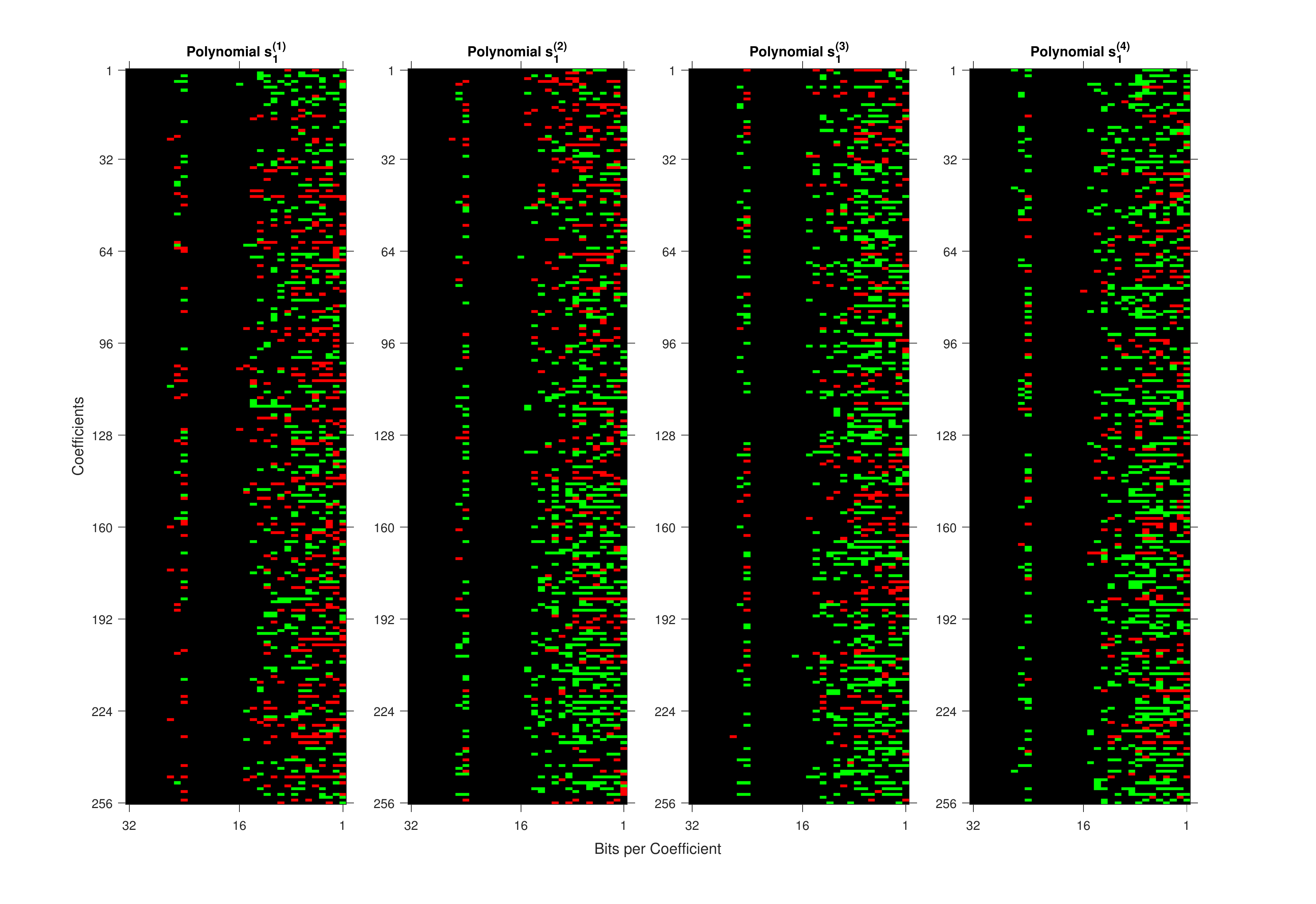}
  \caption{Recovered bits of secret key $s_{1}$ for Dilithium (security level 2). 3,735 in total with 2,454 $0$'s (green pixels) and 1,281 $1$'s (red pixels).}
  \label{fig:recovered_bits}
\end{figure*}

\section{Estimating the Diminished Security Level of Dilithium}
\label{sec:complexity}

\subsection{Lattice Security with Reduced Dimension}
The \attackName Attack can be used iteratively to recover the secret key-bits. There are however two caveats in applying \attackName in practice:
\begin{itemize}
    \item Each \attackName recovers only one secret key bit. For full-key recovery we need at least $1024\times 3$ unique faulty signatures which is rather time-consuming.
    \item As described below in practice we inject faults using Rowhammer, which prevents precise targeting of bits. Thus, we need many more \attackName iterations (and time consuming page re-allocations) in practice.
\end{itemize}

To overcome both problems, we instead opt to recover only a fraction of the key-bits to diminish the security level of Dilithium to a point where the remaining key bits can be recovered using lattice attacks.

Here we estimate the new security level of Dilithium by exploiting the recovered bits by \attackName attack. Briefly, Dilithium is based on the hardness of the MLWE and MSIS problems under the Strong Unforgeability under Chosen Message Attack (SUF-CMA) model. We follow the cost estimation approach of \cite{alkim2016post, ducas2018crystals}, i.e., the MLWE problem is analyzed as an LWE problem and the security level is estimated using standard lattice hardness estimation techniques. Specifically, we base our estimate on the so-called primal and dual attacks \cite{10.1007/978-3-642-25385-0_1,10.1007/BF01581144} and use BKZ for lattice reduction. Cost estimation of the attacks are given in \cite{alkim2016post, ducas2018crystals}. Note that these estimates ignore SVP oracle calls. Instead, core-SVP hardness which is the cost of one call to an SVP oracle in dimension $b$ is taken into account. For Quantum attacks, the Sieve algorithm is used to estimate the core hardness of underlying open problem. For the quantum sieve algorithm the heuristic complexity is $\sqrt{3/2}^{b+O(b)}\approx 2^{0.292b}$ \cite{10.5555/2884435.2884437,10.1007/978-3-662-47989-6_1}. Grover’s quantum search algorithm reduces the complexity down to $2^{0.265b}$ \cite{laarhoven2016search,10.1007/s10623-015-0067-5}. Cost of solving SVP in classical attack bound is $2^{0.2075b}\approx 2^{39}$ for the best-known algorithm \cite{10.1007/978-3-642-25385-0_1}.

The \attackName attack allows us to recover certain number of bits of the private key $s_1$. By analyzing the distribution of the recovered bits, we can recover additional bits. We follow the general methodology from \cite{alkim2016post, bos2016frodo} to analyze the reduced security of Dilithium with side information recovered using the \attackName attack as described in Section \ref{sec:attack}. The reduced security levels can be determined by following the analysis in \cite{ducas2018crystals, alkim2016post, bos2016frodo}. The analysis converts the equation system into an LWE instance of dimensions $256\cdot l$ and $256\cdot k$ by taking the coefficients of polynomial elements in the MLWE problem as the vectors of coefficients in LWE problem. Hence, the problem is reduced to finding the coefficient vectors 
\[
s_1 ~,~ s_2 \in \left(\mathbb{Z}^{256\cdot l} \times \mathbb{Z}^{256\cdot k} \right)
\]
from $\bar{A}$ and coefficient vector of $t$. Here  $\bar{A}\in \mathbb{Z}_q^{256\cdot k\times256\cdot l}$ is obtained by replacing all entries $a_{ij}\in R_q$ of $A$ by the rotation of the coefficient vectors of $a_{ij}$. One can show that the private key coefficients recovered using the \attackName Attack can be used to reduce the dimension $n$ of the lattice formed by the equation system 
\begin{equation}\label{LWE}
As_1+s_2=t 
\end{equation}
by inserting the recovered coefficients of the secret key into its polynomial form. Moreover, inserting the recovered bits which are not used to find the coefficients can also reduce the norm of the coefficient vector. The security estimates for the scheme reduced for a given number of coefficients recovered using \attackName is given in Table~ \ref{tab:RedComplexities}. From the table, we deduce that recovering 8 coefficients of secret key reduces the attack complexity to around half of the overall complexity on average. In other words, recovering approximately 320 coefficients by \attackName attack is enough to reduce the attack complexity to a practical level, i.e. 80 bits. Note that, we take the norm of secret key coefficients as $\zeta=\sqrt{10}$. Estimated time to recover is given in Section \ref{sec:key_recovery}. \footnote{The script ``scripts/PQsecurity.py'' which estimates the cost of primal and dual attacks can be found at \cite{alkim2016post}.} 

To recover 320 coefficients, we need 960 bits in the same coefficients. Therefore, we cannot conclude that any 960-bit recovery is enough to break the scheme since we do not have any control on the location of the recovered bits.

\subsection{Exploiting the Redundant Encoding to Recover More Coefficients}
\label{sec:BitToCoeff}
In this section, we focus on how we can use the recovered bits in the most effective way to diminish the security level of Dilithium. For this purpose, we divide the coefficients of the secret key polynomial into 3 groups: 
\begin{itemize}
    \item {\bf Group 1} has the fully recovered coefficients, i.e, 3 out of 3 bits are known in the coefficients. Number of recovered coefficients in Group 1 can directly be used to {\bf reduce the dimension} $n$ of the LWE system (Section~\ref{sec:BitToCoeff}). 
    \item {\bf Group 2} consists the coefficients in which 1 or 2 out of 3 bits are known in each coefficient. The recovered bits in this category fall short in reducing the LWE dimension further, yet they can still be used to {\bf reduce the norm} of the secret key coefficients, i.e, unique SVP solution in LWE system (Section~\ref{sec:BitToNorm}). 
    \item {\bf Group 3} is the collection of coefficients with no recovered bits.
    hence yielding no information about the coefficients.
\end{itemize}

\begin{table}[]
\centering
\caption{Recovering an additional bit by using recovered 2-bit info by Rowhammer. Shaded rows has the additional bit recovery, i.e., full coefficient is recovered by 2-bit info.}
\label{2bitinfo}
\begin{tabular}{l c c}
\toprule
Known bits & Possible  & $\#$ of possible \\ 
of $xyz$ & coefficients & coefficients \\ \hline
$00z$ & $00z$ & 2 \\ 
\rowcolor{lightgray} $01z$ & $010$ & 1 \\ 
$10z$ & N/A & 0 \\ 
$11z$ & $11z$ & 2 \\ \hline 
$0y0$ & $0y0$ & 2   \\ 
\rowcolor{lightgray} $0y1$ & $001$ & 1 \\ 
\rowcolor{lightgray} $1y0$ & $110$ & 1 \\ 
\rowcolor{lightgray} $1y1$ & $111$ & 1 \\ \hline 
\rowcolor{lightgray} $x00$ & $000$ & 1   \\ 
\rowcolor{lightgray} $x01$ & $001$ & 1 \\ 
$x10$ & $x10$ & 2 \\ 
\rowcolor{lightgray} $x11$ & $111$ & 1 \\     
\bottomrule
\end{tabular}
\end{table}

\begin{table}[]
\centering
\caption{Number of additional full coefficient recoveries by 2-bit info. Highlighted bit shows the additional bit recovery.}
\label{table:extracoeff}
\begin{tabular}{lcccccccc}
\toprule
xyz &  Rec Coeffs   & $s_1^{(1)}$ & $s_1^{(2)}$ & $s_1^{(3)}$ & $s_1^{(4)}$ & Total \\ \hline
01{\bf z} & 01{\bf 0} & 10 & 11 & 6 & 5 & 32 \\ \hline
0{\bf y}1 & 0{\bf 0}1 & 12 & 5 & 6 & 8 & 31 \\
1{\bf y}0 & 1{\bf 1}0 & 7 & 7 & 6 & 9 & 29 \\
1{\bf y}1 & 1{\bf 1}1 & 10 & 9 & 5 & 11 & 35 \\ \hline
{\bf x}00 & {\bf 0}00 & 1 & 0 & 0 & 0 & 1 \\
{\bf x}01 & {\bf 0}01 & 0 & 0 & 0 & 0 & 0 \\
{\bf x}11 & {\bf 1}11 & 0 & 0 & 0 & 1 & 1  \\ \hline
\multicolumn{2}{l}{Total $\#$ of Rec Coeffs}  & {\bf 40} & {\bf 32} & {\bf 23} & {\bf 34} & {\bf 129} \\
\bottomrule
\end{tabular}
\end{table}

When we estimate the security level, our calculations are based on the number of recovered coefficients and the norm of the remaining unknown coefficients. Secret key is defined as an $l$ dimensional vector of $n^{th}$ degree polynomials with coefficients in the range $[-\eta ,\eta ]$. In our experiments, we consider Dilithium security level 2 in which parameters are set to $\eta =2$, $l=4$ and $n=256$ \cite{ducas2018crystals}, i.e., Each coefficient of the secret key is in $\{ -2,-1,0,1,2\}$ but is encoded in the reference implementation as 32-bit  words $\{ 1\cdots 1110,1\cdots 1111,0\cdots 0000,0\cdots 0001,0\cdots 0010\}$, respectively. Therefore we have a highly redundant representation, where the per coefficient entropy of secret key encoding is only 2.25 bits (in 32 logical bits). The recovered bits are distributed over the last three bits of 1024 coefficients given in Figure~\ref{fig:recovered_bits}. Additionally, distribution of number of recovered bits up to the coefficients are given in Table~\ref{Tab:distrecoveredbits}. 

\begin{table}[]
\centering
\caption{Distribution of 1,522 bits recovered by \attackName Algorithm to the \# secret key in polynomial coefficients. Total of 99 coefficients are recovered with another 857 coefficients yielding only partial information.}
\label{Tab:distrecoveredbits}
\begin{tabular}{l|cccc|c|c}
\toprule
& $s_1^{(1)}$ & $s_1^{(2)}$ & $s_1^{(3)}$ & $s_1^{(4)}$ & $\mathbf{\# bits}$ & $\mathbf{\# coefs}$  \\ \hline
No recovery & 19 & 14 & 18 & 17 & 0 & $\mathbf{68}$ \\ 
1 bit rec & 122 & 126 & 131 & 110 & 489 & $\mathbf{489}$ \\ 
2 bits rec & 95 & 89 & 86 & 98 & 736 &$\mathbf{368}$ \\ 
Full rec & 20 & 27 & 21 & 31 & 297 & $\mathbf{99}$ \\ \hline
$\mathbf{Total}$ & $\mathbf{372}$ & $\mathbf{385}$ & $\mathbf{366}$ & $\mathbf{399}$ & $\mathbf{1522}$ & $\mathbf{1024}$\\     
\bottomrule
\end{tabular}
\end{table}

Even though the recovered 1,522 bits are expected to give us information for about 507 coefficients, just 99 coefficients fully recovered, since only 297 out of 1,522 bits are concentrated in 99 coefficients. The remaining 1,423 bits are distributed over the remaining 857 different coefficients. On the other hand, 2-bit recovered in any coefficient yields either 0 or 1 bits of information on a coefficient as summarized in Table~\ref{2bitinfo}. Here coefficients are represented by $xyz$ where $z$ denotes the least significant bit (LSB) of the coefficient, and $x$ represents the the most significant bit (MSB) if we represent the coefficients by the last three bits. All higher order bits will be identical to $x$, i.e. the sign bit of the coefficient. In certain cases, with a 2-bit information of a coefficient we can recover the full 3-bit coefficient as shown in Table~\ref{2bitinfo}. For instance, if we recovered the first two bits as in the case of $01z$, then due to the encoding the only possible value $z$ can take is $0$. We can fully recover a coefficient from 2-bits of information in 7 out of 12 cases as shown in the shaded rows in Table~\ref{2bitinfo}. With this approach, we managed to recover an additional 129 coefficients of the secret key as summarized in Table~\ref{table:extracoeff}. The total number of recovered coefficients is increased significantly, i.e. from 99 to 228. You can find the number of recovered coefficients in Table~\ref{recoveredbits}.

\begin{table}[]
\centering
\caption{Recovering an additional bit by using 1-bit recovered by Rowhammer. Shaded rows yield an extra bit.}
\label{1bitinfo}
\begin{tabular}{l c c}
\toprule
Known bit & Possible  & $\#$ of possible  \\ 
of $xyz$ & coefficients & coefficients \\ \hline
\rowcolor{lightgray} $1yz$ & $11z$ & 2   \\ 
$0yz$ & $00z$ or 010 & 3 \\ \hline
$x1z$ & $11z$ or 010          & 3 \\ 
\rowcolor{lightgray} $x0z$ & $00z$ & 2   \\ \hline
\rowcolor{lightgray} $xy1$ & 001 or 111 & 2   \\ 
$xy0$ & $x10$ or 000 & 3 \\     \bottomrule
\end{tabular}
\end{table}

\begin{table}[]
\centering
\caption{Number of additional bit recovery by 1-bit info. Highlighted bit shows the recovered bit.}
\label{table:extrabit}
\begin{tabular}{lcccccccc}
\toprule
xyz &  Rec Coeffs   & $s_1^{(1)}$ & $s_1^{(2)}$ & $s_1^{(3)}$ & $s_1^{(4)}$ & Total \\ \hline
1yz & 1{\bf 1}z & 51 & 46 & 56 & 37 & 190 \\ 
x0z & {\bf 0}0z & 2 & 1 & 2 & 0 & 5 \\
xy1 & {\bf xx}1 & 2 & 1 & 1 & 1 & 5 \\ \hline
\multicolumn{2}{l}{Total $\#$ of Rec bits}  & {\bf 55} & {\bf 48} & {\bf 59} & {\bf 38} & {\bf 200} \\
\bottomrule
\end{tabular}
\end{table}

\subsection{Reducing the Norm of the Coefficients}
\label{sec:BitToNorm}
In cases where we recover 1-bit out of a coefficient the information is not sufficient to recover the entire coefficient. However, we can still gain information useful in reducing the attack complexity. Specifically we can reduce the norm of the target vector by removing known bits from it. This reduces the complexity of the lattice search problem. 

Further in certain cases the 1-bit knowledge may facilitate recovery of an additional bit of the coefficient. In Table~\ref{1bitinfo}, these special cases are shown in shaded rows. Analyzing the experimentally recovered bits gives us 200 of these special cases, i.e., two bits of 200 coefficients are recovered by 1-bit information. In Table~\ref{table:extrabit}, the number of coefficients in which extra bit recovery is possible is shown. By analyzing the recovered bits, we recovered 1 bit of 289 coefficients and 2 bits of 439 coefficients. There are 68 coefficients that we have no extra information about. Number of recovered bits and coefficients by Rowhammer and extra bit recovery method is given in Table~\ref{recoveredbits}. When we insert these recovered bits into the Lattice formulation the norm of secret key coefficients in the reduced system is decreased to 
\[
    \zeta = \frac{68}{796}\times 3+\frac{289}{796}\times 2+\frac{439}{796}\times 1=1.53392.
\]

By analyzing the encoding (Section \ref{sec:key_recovery}), we increased the number of recovered bits from 883 to 1,522. This was achieved by taking recovered bits from 4 to 32 as the sign bit, i.e. $x$. Then we further increased from 1,522 to 1,851 by considering the positions in the recovered bits in the last 3 bits of the coefficient. In total, the number of fully recovered coefficients are increased from 99 to 228, and the number of coefficients with 2 bits known are increased from 368 to 439. By analyzing the encoding, we partially or fully recovered 956 coefficients of 1024 secret key coefficients, in total. The breakdown is given in Table~\ref{recoveredbits}. 

The diminished security level of Dilithium with the fully recovered coefficients (reduced dimension $\bar{n}$) and reduced average norm $\zeta$ is listed in Table~\ref{tab:RedComplexities}. Note that with the fully recovered coefficients the reduced security level is 124-bits for classical and 112-bits for quantum attackers. By also exploiting the encoding to increase the fully recovered coefficients from 99 to 228 and partially recovered coefficients to reduce the norm from $\zeta=\sqrt{10}$ to $\zeta=1.53392$, we managed to significantly degrade the security level: {\bf 89-bits (classical) and 81-bits (quantum)}.

\begin{table*}[]
\centering
\caption{Number of Recovered Information by \attackName up to the number of coefficients. Highlighted rows show the number of coefficients with additional bit recovery.}
\label{recoveredbits}
\begin{tabular}{l|cccc|c}
\toprule
& $s_1^{(1)}$ & $s_1^{(2)}$ & $s_1^{(3)}$ & $s_1^{(4)}$ & $\mathbf{\# coefs}$ \\ 
\hline 
\hline
\multicolumn{6}{c}{Group 3: Coefficients with \textbf{no bit} recovery.} \\ 
\multicolumn{6}{c}{\textbf{68} coefficients in total.} \\
\hline
No recovery & 19 & 14 & 18 & 17 & $\mathbf{68}$ \\  
\hline
\hline
\multicolumn{6}{c}{Group 2: Coefficients with \textbf{1} bit recovery. } \\ 
\multicolumn{6}{c}{\textbf{289} bits in \textbf{289} coefficients in total.} \\ 
\hline
1 bit rec by 1 bit & 67 & 78 & 72 & 72 & $\mathbf{289}$  \\ 
\hline
\multicolumn{6}{c}{Group 2: Coefficients with \textbf{2} bit recovery.} \\ 
\multicolumn{6}{c}{\textbf{878} bits in \textbf{439} coefficients in total.} \\ 
\hline
\rowcolor{lightgray}2 bits rec by 1 bit & 55 & 48 & 59 & 38 & $\mathbf{200}$ \\ %\cline{1-6}
2 bits rec by 2 bits & 55 & 57 & 63 & 64 & $\mathbf{239}$  \\ 
\hline
\hline
\multicolumn{6}{c}{Group 1: Full coefficient recovery. } \\
\multicolumn{6}{c}{\textbf{684} bits in \textbf{228} coefficients in total.} \\ \hline
\rowcolor{lightgray}Full Coef rec by 2 bits & 40 & 32 & 23 & 34 & $\mathbf{129}$ \\ 
Full Coefs rec by 3 bits & 20 & 27 & 21 & 31 & $\mathbf{99}$ \\ 
\hline
\hline
$\mathbf{Total \# rec bits (\textbf{1851})}$ & $\mathbf{467}$ & $\mathbf{465}$ & $\mathbf{448}$ & $\mathbf{471}$ & $\mathbf{1024}$  \\     
\bottomrule
\end{tabular}
\end{table*}

\begin{table*}
\centering
    \caption{The reduced security level of Dilithium using the \attackName Attack. The value $\bar{n}$ denotes the reduced lattice dimension, $b$ the block dimension of BKZ, and $m$ the number of samples. Cost is given in log base 2 and is the smallest cost for all possible choices of $m$ and $b$. Shaded rows show improvements: 124-bits (classical) and 112-bits (quantum) with plain \attackName, 89-bits (classical) and 81-bits (quantum) by also exploiting the encoding in addition to \attackName .
    }
    \label{tab:RedComplexities}
	\begin{tabular}{cc||c||llll||llll}
		\toprule
		\multicolumn{11}{l}{Dilithium Security Level II (128 bit) parameters: $q=2^{23}-2^{13}+1$, $n=1024$} \\
		\hline
        \hline
        & & & & & Primal Attack & & & & Dual Attack & \\

		$\#$Rec coeffs& $\bar{n}$ & $\zeta$ & m & b & Classical & Quantum & m & b & Classical & Quantum \\
		\hline
		\hline
		0 & 1024 & $\zeta=\sqrt{10}$ & 1090 & 485 & 141 & 128 & 1089 & 484 & 141 & 128 \\ \hline
        0 & 1024 & $\zeta=1.53392$ & 1001 & 429 & 125 & 113 & 1027 & 428 & 125 & 113 \\
		\hline
        \multicolumn{11}{c}{Reduced Complexities with $\#$ Recovered coefficients and Reduced Norm} \\
		\hline
		\hline
		1 & 1023 & $\zeta=\sqrt{10}$ & 1129 & 484 & 141 & 128 & 1132 & 483 & 141 & 128 \\
		2 & 1022 & $\zeta=\sqrt{10}$ & 1075 & 484 & 141 & 128 & 1074 & 483 & 141 & 128 \\
		4 & 1020 & $\zeta=\sqrt{10}$ & 1062 & 483 & 141 & 128 & 1062 & 482 & 140 & 127 \\
		8 & 1016 & $\zeta=\sqrt{10}$ & 1089 & 480 & 140 & 127 & 1090 & 479 & 140 & 127 \\
        64 & 960 & $\zeta=\sqrt{10}$ & 1025 & 446 & 130 & 118 & 1037 & 445 & 130 & 118 \\
		\rowcolor{lightgray}99 & 925 & $\zeta=\sqrt{10}$ & 981 & 425 & 124 & 112 & 997 & 424 & 124 & 112 \\
		128 & 896 & $\zeta=\sqrt{10}$  & 933 & 408 & 119 & 108 & 947 & 407 & 119 & 107 \\
		192 & 832 & $\zeta=\sqrt{10}$ & 919 & 369 & 107 & 97 & 885 & 369 & 107 & 97 \\
		228 & 796 & $\zeta=\sqrt{10}$ & 863 & 348 & 101 & 92 & 843 & 348 & 101 & 92 \\
		288 & 736 & $\zeta=\sqrt{10}$ & 799 & 313 & 91 & 83 & 788 & 313 & 91 & 83 \\
		320 & 704 & $\zeta=\sqrt{10}$ & 744 & 295 & 86 & 78 & 810 & 294 & 86 & 78 \\
		352 & 672 & $\zeta=\sqrt{10}$ & 745 & 276 & 80 & 73 & 742 & 276 & 80 & 73 \\ 
		\hline
			99 & 925 & $\zeta=1.53392$ & 902 & 375 & 109 & 99 & 957 & 374 & 109 & 99 \\
		\rowcolor{lightgray}228 & 796 & $\zeta=1.53392$ & 782 & 306 & 89 & 81 & 773 & 306 & 89 & 81 \\
		\bottomrule
	\end{tabular}
\end{table*}

%%%%%%%%%%%%%%%%%%%%%%%%%%%%%%%%%%%%%%%%%%%%%%%%
\section{Discussion}
\subsection{Is the weakness inherent to Dilithium?}
In our attack we exploited the linear structure of Step~\ref{lst:line:z} in the Dilithium Signing Algorithm:
$$z \gets y+c\cdot s_{1}~.$$
To this end, we compute and check possible fault patterns as they would appear as additive terms in the faulty signature $\bar{z}$. This approach is enabled by the {\em linearly additive} secret mask $y$ and the publicly known challenge vector $c$. Clearly, the presented signature correction algorithm is specific to Dilithium. However, we have also tried to produce a similar technique in the GeMSS \cite{casanova2017gemss} and Rainbow \cite{ding2005rainbow} schemes which gave insufficient partial information. While the approach is generic, the particulars of the signing algorithm may still make it hard to trace the fault to the output without causing the search space to grow exponentially, thus preventing efficient signature correction. While the presented attack utilizes faulty signatures to recover secret key bits we have also exploited the highly redundant encoding of the coefficients to gain significant advantage in reducing the security level of Dilithium. This weakness is not rooted in the algorithm itself, but rather due to the choice of representation used in the implementation.

\subsection{Further Reducing the Attack Complexity}
Dachman-Soled et al. \cite{dachman2020lwe} introduced a framework for cryptanalysis of lattice based schemes when side-information in the form of ``hints'' about the secret and/or error is available. The framework allows the primal lattice reduction attack and allows  progressive integration of hints before running a lattice reduction step. What we refer to as ``recovered coefficient'' and ``partially recovered coefficient'' correspond to ``Modular hints'' and ``Approximate hints'', respectively. Along with the framework the authors introduced techniques for progressively sparsifying the lattice, projecting onto and intersecting with hyperplanes, and/or altering the distribution of the secret vector. One may apply these more advanced techniques to gain advantage and further degrade the security level.

\section{Countermeasures}
\label{sec:countermeasures}

Every novel attack sheds light onto how to strengthen a cryptographic scheme, and in this perspective, a discussion on countermeasures is very important. We can find considerable work on countermeasures against fault attacks on PQC schemes \cite{bruinderink2018differential, ravi2019exploiting, espitau2016loop, bindel2016lattice}. In particular, Bindel {\em et al.} \cite{bindel2017special} have  written an exhaustive literature review on countermeasures for fault attacks on lattice-based signature schemes.

For our \attackName attack, there are two ways to detect and prevent the fault attack. First, we can prevent or detect the fault injection mechanism, which means that we would prevent or detect Rowhammer faults. Second, we can prevent or detect the exploitation of an injected fault. This requires an algorithmic countermeasure, such as preventing faulty signatures from being returned by the signer. Algorithmic countermeasures are required because our attack is independent of the fault mechanism used. In the following, we discuss the Rowhammer countermeasures followed by algorithmic countermeasures. Then, we provide a literature review of countermeasures against implementation attacks on lattice-based signature and encryption schemes in Table \ref{tab:countermeasures}. In this table, we have shown countermeasures which work against timing, cache and fault attacks with a green tick mark and which doesn't work with a red cross mark. We show that post-quantum schemes are broadly vulnerable to three kinds of fault attacks, DFA, Instruction Skip and single-bit trace analysis. The table describes how countermeasures help against these known attacks, which include an attack on an older round-2 PQ scheme \cite{mus2020quantumhammer} as well as our proposed \attackName.

\newcommand{\cmark}{\ding{51}}
\newcommand{\xmark}{\ding{55}}
\renewcommand\arraystretch{1.1}
\newcommand{\tikzcirclered}[2][red,fill=red]{\tikz[baseline=-0.5ex]\draw[#1,radius=#2] (0,0) circle ;}%
\newcommand{\tikzcirclegreen}[2][green,fill=green]{\tikz[baseline=-0.5ex]\draw[#1,radius=#2] (0,0) circle ;}%

\newcommand{\ok}{\textcolor{green}{\large{\cmark}}}
\newcommand{\bad}{\textcolor{red}{\large{\xmark}}}

\begin{table*}
\centering
\caption{{An Overview of Countermeasures against Implementation Attacks on Lattice-Based Post-Quantum Cryptography; \ok Countermeasure works, \bad Countermeasure doesn't work}}
\label{tab:countermeasures}
\begin{tabular}{ccccccc}
    \toprule
    \multirow{8.0}{*}{\parbox[t][][c]{2.5cm}{\centering \textbf{Countermeasures}}} & \multicolumn{6}{c}{\parbox[t][][c]{3.0cm}{\centering \textbf{Implementation Attacks}}}\\[11pt]
    \cline{2-7}
    {} & \multirow{6.5}{*}{\parbox[t][][c]{0.5cm}{\centering \textbf{Timing\\}\cite{8226020, 10.1007/11967668_14, 10.1145/3338467.3358948}}} & \multirow{7.5}{*}{\parbox[t][][c]{1.0cm}{\centering \textbf{Cache\\} \cite{bruinderink2016flush, pessl2017bliss, 8226020, bindel2017bounding}}} & \multicolumn{4}{c}{\multirow{1.5}{*}{\parbox[t][][c]{7.5cm}{\centering \textbf{Fault}}}}\\[10pt]
    \cline{4-7}
    {} & {} & {} & \multirow{4.0}{*}{\parbox[t][][c]{1.5cm}{\centering \textbf{DFA\\}\cite{castelnovi2018grafting, bruinderink2018differential}}} & \multirow{4.0}{*}{\parbox[t][][c]{2.0cm}{\centering \textbf{Instruction Skip\\}\cite{ravi2019number, bruinderink2018differential, ravi2019exploiting, bindel2016lattice, espitau2016loop, XagawaIUTH21, 10.1145/3178291.3178294, NTRUinstskip}}} & \multirow{4.0}{*}{\parbox[t][][c]{2.0cm}{\centering \textbf{QuantumHammer} \cite{mus2020quantumhammer} \\
    (LUOV Round2)}} & \multirow{4.0}{*}{\parbox[t][][c]{2.0cm}{\centering \textbf{Signature Correction Attack}\\ (this work)}} \\[35pt]
    \midrule
    \textbf{Constant run-time \&} \\ \textbf{data-oblivious accesses} \cite{bruinderink2016flush} & \ok & \ok  & \bad & \bad & \bad & \bad\\
    \textbf{Key-independent control flow} \\ \textbf{\& memory accesses} \cite{8226020} & \bad & \ok & \bad & \bad & \bad & \bad\\
    \textbf{Nonce Randomization} \cite{bruinderink2018differential, ravi2019exploiting} & \bad & \bad & \ok & \ok & \bad & \bad\\
    \textbf{Temporal Redundancy} \cite{castelnovi2018grafting, ravi2019exploiting, Kamal2013StrengtheningHI} & \bad & \bad & \ok & \ok & \bad & \bad\\
    \textbf{Spatial Redundancy} \cite{castelnovi2018grafting, Kamal2013StrengtheningHI}& \bad & \bad & \ok & \bad & \ok & \ok\\ 
    \textbf{Verify-after-sign} \cite{castelnovi2018grafting, ravi2019exploiting} & \bad & \bad & \ok & \ok & \ok & \ok\\
    \textbf{HPC} \cite{gulmezoglu2019fortuneteller} & \bad & \bad  & \bad & \bad & \ok & \ok\\
    \textbf{DRAM Refresh Rate} \cite{mutlu2019rowhammer} & \bad & \bad & \bad & \bad & \ok & \ok\\
    \bottomrule
\end{tabular}
\end{table*}

\subsection{Rowhammer Countermeasures}

We discuss two approaches to counter Rowhammer attack. One technique detects the Rowhammer attack through hardware monitors, while the second technique prevents Rowhammer from happening in the first place.

\paragraph*{\textbf{Hardware Performance Counters (HPC)}}
HPCs are special purpose registers which store the hardware events inside the CPU like cache hits and cache missed. As the Rowhammer bypasses the cache and directly hits the DRAM, there will be a significant increase in the number of cache misses which can be used to detect the Rowhammer attack. These HPCs, when paired with machine learning techniques, can detect Rowhammer attack with high accuracy. Gulmezoglu {\em et al.} \cite{gulmezoglu2019fortuneteller} have shown an accuracy of 100\% using the performance counter \emph{LLC\_Miss}.

\paragraph*{\textbf{Increasing DRAM Refresh Rate}}
DDR3 and DDR4 specifications require that each DRAM row must be refreshed after at least 64ms to retain its values \cite{standard2005double}. However, as we have seen this refresh rate is not sufficient in Rowhammer scenarios where hammering the neighboring rows cause the cells to leak faster than normal and are unable to retain their charge. So, an immediate mitigation can be to decrease the refresh interval to 32ms or 16ms. Many systems allow this configuration from the BIOS for better memory stability. However, there are two downsides for this approach. The first one is that the power consumption will increase and the second one is the reduction of data transfer rate. This is because while the cells are refreshed, the data can not be read or written. Also, the Rowhammer can not be fully mitigated by this countermeasure. At most, one can significantly reduce the chances of getting bit flips. Mutlu {\em et al.} \cite{mutlu2019rowhammer} have shown that to fully mitigate Rowhammer using the refresh rate, one needs to set the refresh rate as 8.2 ms which is 7.8 times lower than 64 ms. This will cause significant burden on power consumption and quality of service which researches are already trying to improve \cite{chang2014improving, liu2012raidr}.

\subsection{Algorithmic Countermeasures}
Here, we discuss algorithmic countermeasures for PQC signature schemes, specifically Dilithium against general fault attacks as well as our \attackName attack. These countermeasures include adding randomization, temporal and spatial redundancy techniques and verify-after-sign approach.

\paragraph*{\textbf{Randomized Signing}}
Due to the fault attacks based on determinism like \cite{bruinderink2018differential, ravi2019exploiting}, Dilithium added this mitigation in Round 2 for DFAs as listed in step \ref{lst:line:randomized} of Algorithm \ref{alg:dilithium_sign}. Here, the nonce $y$ is generated randomly instead of generating using the message $M$, recommended for side-channel and fault attacks \cite{ducas2018crystals}. The idea is that if the attacker can not collect the faulty and correct signature pair for the same message, the DFA will not work. However, this mitigation will not work for our \attackName attack as it is independent of the nonce $y$. Whatever the value of the $y$ is, we get the same error in the faulty signature depending upon the position of the fault in secret key $s_1$. 

\paragraph*{\textbf{Temporal Redundancy}}
Temporal Redundancy requires re-execution of same task after a certain amount of time and comparing their results. If they don't match then the output of the algorithm is disabled in order to prevent the attacker from accessing any information from the faulty signature. It makes harder for the attacker to inject the same fault in redundant computations.  However, as the Rowhammer attack faults the memory and can induce permanent faults, if multiple signatures are generated using the same faulty secret key in memory, they will still match, fault remains undetected. An algorithm with such a countermeasure can provide fault tolerance against transient faults but not permanent faults. Hence, it is recommended to add spatial redundancy.

\paragraph*{\textbf{Spatial Redundancy}}
Spatial Redundancy involves simultaneous execution of $N$ instances of a critical task for $N$ level of redundancy and comparing their results for fault detection. If we store redundant copies of the original secret key during the key generation process at different memory locations, they can be used in parallel computations of signature generation using spatial redundancy technique. To bypass this approach, the attacker will need to fault the same exact bits at both memory locations which will be much harder because every cell in the DRAM is not faulty. An important point here is that for deterministic version of Dilithium, this approach can work in a straightforward manner because the same nonce $y$ is generated for the same message signed twice. However, for the randomized version, we will also need to store a copy of the nonce $y$ for redundant computation. On the performance side, computation based on spatial redundancy will have significant overhead than the normal computation because of the increased complexity of the algorithm. Therefore, the level of spatial redundancy needed to detect faults should be taken into consideration. 

\paragraph*{\textbf{Verify-after-Sign}}
If there is any bit flip in the secret key $s_1$ of Dilithium, it will generate a faulty signature which will not be verified by the verification algorithm. Hence, the verification can be used as a fault detection mechanism and if done on the signing side, the sender can easily detect the existence of the attacker. As compared to double signing, this approach is approximately three times faster as the verification algorithm takes less time as compared to the signing algorithm. There is still a possibility that the attacker also faults the verification in a way which results in a valid signature but to the best of our knowledge, there has not been such an algorithm developed so far for Dilithium. This approach may also fail if the comparison instruction is skipped using an instruction skip fault similar to \cite{ravi2019exploiting}. Verify-after-sign can also detect instruction skip faults on signing step \ref{lst:line:z} and the rejection sampling step \ref{lst:line:rejection_sampling} in Algorithm \ref{alg:dilithium_sign} \cite{ravi2019exploiting}. Rejection sampling step is critical because if it is bypassed without a mitigation in place, it can reveal information about the secret key \cite{ducas2018crystals}.

\section{Conclusion}
\label{sec:conclusion}
We have proposed the \attackName attack targeting Dilithium a Round 3 finalist in the NIST PQC competition. The attack requires single bit-flips in the secret key vector, which we have implemented using Rowhammer targeting the constant-time AVX2 reference implementation of Dilithium. By analyzing the faulty signatures and exploiting redundancy in the secret key encoding, our attack successfully recovered 1,851 bits (out of 3,072 bits) of the secret key. This enabled us to reduce the post quantum security level to 81-bits (quantum) and 89-bits (classical) for both primal and dual lattice attacks. The attack is also applicable to other variants and security levels. We have demonstrated the first fault attack which works on randomized as well as deterministic versions of Dilithium. Our \attackName attack is independent of the fault mechanism but we have used Rowhammer to demonstrate the attack as it is a software only attack and does not need physical access. This can be very critical in case of cloud scenarios where the attacker can launch an attack remotely and leak secret information by using only faulty signatures. We have shown how few bits of secret key significantly reduce the security strength of Dilithium using the lattice attacks especially when the secret key encoding is exploited as in the analysis shown in this paper. Further, randomizing the nonce, a countermeasure commonly implemented in the design of signature schemes to thwart fault attacks, is not sufficient in practice as demonstrated by our attack.

% conference papers do not normally have an appendix

% use section* for acknowledgment
\section*{Acknowledgment}
We thank our anonymous reviewers for their insightful comments for improving the quality of this paper. This work is supported by U.S. Department of State, Bureau of Educational and Cultural Affair's Fulbright Program and by the National Science Foundation under grants CNS-1814406, CNS-2026913 and CNS-1931639.

%\newpage

% trigger a \newpage just before the given reference
% number - used to balance the columns on the last page
% adjust value as needed - may need to be readjusted if
% the document is modified later
\IEEEtriggeratref{94}
% The "triggered" command can be changed if desired:
%\IEEEtriggercmd{\enlargethispage{-5in}}

% references section

% can use a bibliography generated by BibTeX as a .bbl file
% BibTeX documentation can be easily obtained at:
% http://mirror.ctan.org/biblio/bibtex/contrib/doc/
% The IEEEtran BibTeX style support page is at:
% http://www.michaelshell.org/tex/ieeetran/bibtex/
\bibliographystyle{IEEEtran}
% argument is your BibTeX string definitions and bibliography database(s)
\bibliography{references}
%
% <OR> manually copy in the resultant .bbl file
% set second argument of \begin to the number of references
% (used to reserve space for the reference number labels box)
%\begin{thebibliography}{1}

%\bibitem{IEEEhowto:kopka}
%H.~Kopka and P.~W. Daly, \emph{A Guide to \LaTeX}, 3rd~ed.\hskip 1em plus
%  0.5em minus 0.4em\relax Harlow, England: Addison-Wesley, 1999.

%\end{thebibliography}

% that's all folks
\end{document}